\documentclass[11 pt]{article}
\usepackage{graphicx}
\title{ Fast-neutron induced background in LaBr$_3$:Ce detectors}

\author{J. Kiener$^{1,*}$ \and V. Tatischeff$^{1}$ \and I. Deloncle$^{1}$ \and N. de S\'er\'eville$^{2}$ \and P. Laurent$^{3,4}$ \and
C. Blondel$^{5}$ \and M. Chabot$^{2}$ \and R. Chipaux$^{6}$ \and A. Coc$^{1}$  \and  S. Dubos$^{5}$ \and  A. Gostoji\`{c}$^{1}$ \and 
 N. Goutev$^{7}$ \and  C. Hamadache$^{1}$ \and  F. Hammache$^{2}$ \and  B. Horeau$^{5}$ \and  O. Limousin$^{5}$ \and 
 S. Ouichaoui$^{8}$ \and  G. Pr\'evot$^{4}$ \and  R. Rodr\'{i}guez-Gas\'en$^{1,9}$ \and  M. S. Yavahchova$^{7}$}

\begin{document}

\maketitle

\noindent
$^1${Centre de Sciences Nucl\'eaires et de Sciences de la Mati\`ere (CSNSM), CNRS-IN2P3 et Universit\'e Paris-Sud, 91405 Campus Orsay, France}

\noindent
$^2${Institut de Physique Nucl\'eaire d'Orsay, CNRS-IN2P3 and Universit\'e Paris-Sud, 91406 Orsay, France}

\noindent
$^3${CEA/IRFU Service d'Astrophysique, Orme des Merisiers, CEA Saclay,  91191 Gif-sur-Yvette and Laboratoire Astroparticules et Cosmologie (APC),  10, rue A. Domon et L. Duquet, 75205 Paris, France}

\noindent
$^4${Laboratoire Astroparticules et Cosmologie (APC), 10, rue A. Dormon et L. Duquet, 75205 Paris, France}

\noindent
$^5${Laboratoire AIM, CEA/IRFU, Orme des Merisiers, CEA Saclay,  91191 Gif-sur-Yvette, France}

\noindent
$^6${CEA/DMS/IRFU/SEDI, CEA Saclay, 91191 Gif sur Yvette, France}

\noindent
$^7${Institute of Nuclear Research and Nuclear Energy (INRNE), Bulgarian Academy of Sciences, Sofia, Bulgaria}

\noindent
$^8${USTHB, Facult\'e de Physique, BP 32, El-Alia, 16111 Bab-Ezzouar, Algiers, Algeria}

\noindent
$^9${ Laboratoire d'Etudes Spatiales et d'Instrumentation en Astrophysique (LESIA), Observatoire de Paris-Meudon, CNRS, Meudon, France}

\noindent
$*$ Jurgen.Kiener@csnsm.in2p3.fr
\vspace{1 cm} 

Abstract: The response of a  scintillation detector with a cylindrical 1.5-inch LaBr$_3$:Ce crystal  to incident  neutrons has been measured in the energy range $E_n$ = 2-12 MeV.  Neutrons were produced by proton irradiation of a Li target at $E_p$ = 5-14.6 MeV with pulsed proton beams. Using the time-of-flight information between target and detector, energy spectra of the LaBr$_3$:Ce detector resulting from fast neutron interactions have been obtained at 4 different neutron energies. Neutron-induced $\gamma$ rays emitted by the LaBr$_3$:Ce crystal were also measured in a nearby Ge detector at the lowest proton beam energy.  In addition, we obtained data for neutron irradiation of a large-volume high-purity Ge detector and of a NE-213 liquid scintillator detector,  both serving as monitor detectors in the experiment. Monte-Carlo type simulations for neutron interactions in the liquid scintillator, the Ge and LaBr$_3$:Ce crystals have been performed and compared with measured data.  Good agreement being obtained with the data, we present the results of simulations to predict the response of LaBr$_3$:Ce detectors for a range of crystal sizes to neutron irradiation in the energy range $E_n$ = 0.5-10 MeV. 
\vspace{0.5 cm}

Neutron irradiation; LaBr3:Ce; HP-Ge; NE-213; Monte Carlo simulation
\vspace{0.5 cm}

\section{Introduction}

Interactions of fast neutrons in  detectors are an important source of background in high-energy astronomy and in many nuclear physics experiments. Instruments  for $\gamma$-ray astronomy need to be above Earth's atmosphere in stratospheric balloon experiments or on satellites where they are subject to an intense, complex radiation environment generated by the cosmic rays and their secondary's. The interaction of these particles  with the instrument and platform material creates abundant neutrons, which in turn can produce a significant detector background. Particles in Earth's radiation belts and occasional solar energetic particle events may temporarily increase significantly the radiation background. In low-Earth orbits and on stratospheric balloons, secondary neutrons from cosmic-ray interactions with Earth's atmosphere have to be considered as well. Despite being only a secondary component, neutrons may be responsible for a major part of the detector background in space and balloon-borne $\gamma$-ray telescopes because of their large effective stopping ranges. This is in particular true for instruments equipped with active shields that suppress effectively charged-particle induced background by electronic vetoing, while neutrons may easily penetrate the shield without significant energy deposit and escape detection by the anticoincidence system. 

In nuclear physics experiments at accelerator facilities, neutrons are a natural byproduct of the nuclear reactions involved. Their production cross section in nuclear reaction studies is usually comparable to that for $\gamma$ rays, meaning that a significant part of events in  $\gamma$-ray detectors can be due to neutron interactions. This hampers especially the determination of $\gamma$-ray emission cross sections of lines that are weak, broad or close to prominent neutron-induced structures and strongly disturbs the measurement of $\gamma$-continuum emissions. A better knowledge of neutron interactions in important materials used for $\gamma$-ray detectors is therefore desirable. 

We studied the fast neutron irradiation of 3 types of $\gamma$-ray detectors in the present experiment: our investigations were focused on the LaBr$_3$ scintillator crystal, but we obtained also data for a HP-Ge detector and a liquid scintillator detector that both served as monitors during the experiment. 

Detectors based on liquid scintillators, like NE-213,  are routinely used in nuclear physics for neutron detection, their excellent timing properties allowing neutron spectroscopy by the time-of-flight technique. In $\gamma$-ray astronomy it was for example used in the COMPTEL instrument of the CGRO mission \cite{Comptel}. In our experiment, a NE-213 liquid scintillator detector was used for the neutron flux measurements.

The availability of large-volume high-purity Ge crystals (HP-Ge) with its outstanding energy resolution has for example led to its systematic use in  large $\gamma$-ray detector arrays at nuclear physics facilities. Gamma-ray astronomy was opened up to high-resolution spectroscopy since the launch of $\gamma$-ray instruments with Ge detectors onboard the HEAO-3 \cite{HEAO}, WIND \cite{TGRS} , RHESSI \cite{Rhessi} and INTEGRAL \cite{SPI} spacecrafts. 
In the present experiment, 2 HP-Ge detectors were employed during the neutron irradiations, one for the measurement of the $\gamma$-ray emission of a LaBr$_3$ crystal, the other as a $\gamma$-ray background monitor detector.  

The LaBr$_3$ scintillator crystal is currently employed and under study in various nuclear physics and  $\gamma$-ray astronomy instruments. The recently developed cerium-doped lanthanum bromide (LaBr$_3$:Ce) inorganic scintillator is very promising for both fields. (i) It has a high stopping power and can be grown in large volumes, thus offering the possibility of high detection efficiency  for medium-to-high-energy $\gamma$ rays. (ii) Its energy resolution is comparable to the one of semiconductors operating at room temperature. (iii) With a 1/e fluorescent decay time of 16 ns  \cite{StGobain} it is one of the fastest scintillating materials, thus offering the opportunity for  fast-timing purposes and efficient background rejection. Thus, this scintillator has been chosen for the Mercury Gamma and Neutron Spectrometer (MGNS) of the BepiColombo mission \cite{BepiColombo}, and will also be used in the high-energy instrument of the TARANIS mission \cite{TARANIS}, both scheduled to be launched in the near future. In nuclear physics, it is already used for  timing down to the tens-of-ps range in nuclear  decay studies \cite{RCD} and proposed in a european project  of a 4$\pi$ $\gamma$-ray calorimeter at facilities providing radioactive and stable ion beams  \cite{Paris}.  We studied in detail the response of a 1.5-inch LaBr$_3$:Ce detector to fast neutron irradiation in the energy range $E_n$ = $\sim$2-12 MeV.

The experimental setup is described in the next section, followed by the presentation of the data analysis, the obtained results and  a conclusion. 

\section{Experiment}

\subsection{Setup}

\begin{figure}[h]
\includegraphics[width=10 cm]{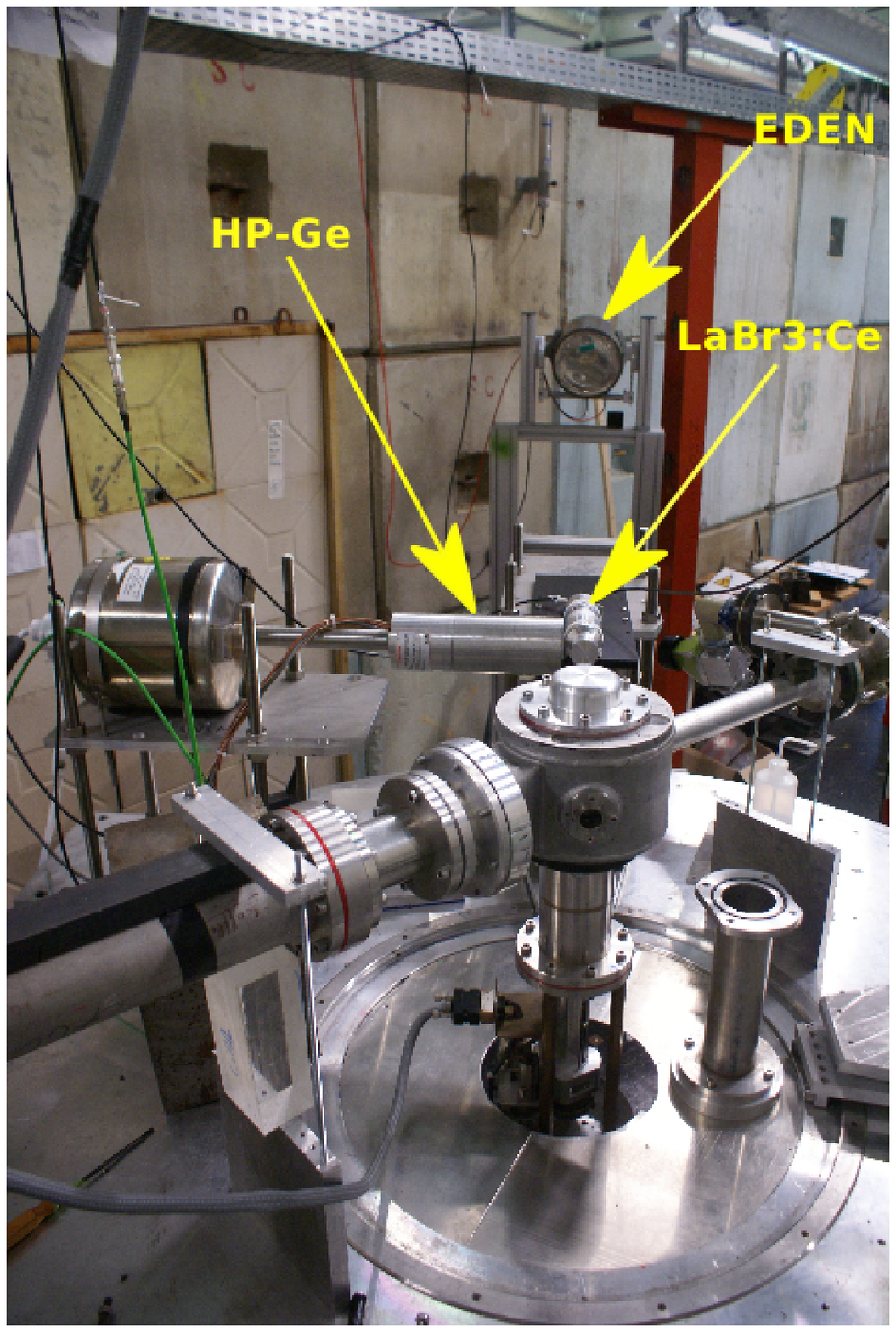}
\caption{  Part of the experimental setup: close to the center is the  target chamber with parts of the pumping station and upstream and downstream beam line (beam is incident from the left), below is the large Al table, behind the LaBr$_3$ pointing to the target position, and the HP-Ge detector pointing to the LaBr$_3$ crystal, and above the EDEN detector. During the irradiation runs paraffin and lead blocks were placed in the space between the HP-Ge detector and the target chamber. Another HP-Ge detector was also placed at $\Theta$ = 135$^{\circ}$ pointing at the target position, approximately at the place of the camera in this photo.   }
\label{photo_setup}
\end{figure}

The experiment was performed at the Tandem-ALTO facility at Orsay. Neutrons were produced by the irradiation of self-supporting Li target foils of 6 and 10 mg/cm$^2$ with pulsed proton beams delivered by the 15-MV tandem Van-de-Graaff accelerator. The pulsation characteristics were a repetition rate of 5 MHz and a pulse width of $\Delta t$ = 1 ns. The Li foils were placed in a small cylindrical vacuum chamber of  10 cm radius and with 2-mm thick Al walls at the center of a large Al table. A 1.5-inch LaBr$_3$ detector  was  fixed onto a  adjustable bench on the large Al table. Its distance from the target was 40 cm and the angle with respect to the beam direction was set to $\Theta$ = 45$^{\circ}$. A large HP-Ge detector (named GV-Orsay) was placed perpendicular to  this detector, very close to the LaBr$_3$ crystal. It was shielded from radiation originating in the target chamber by a 20-cm thick paraffin and 5-cm thick lead block. For the neutron monitoring,  a detector of the EDEN array \cite{EDEN} was mounted on an independent support above the horizontal plane at an angle $\Theta$ = 46$^{\circ}$ with respect to the beam direction at a distance of 223 cm from the target position. These elements and part of the beam line  upstream and downstream of the target chamber can be seen in Fig. \ref{photo_setup}. There was furthermore another large HP-Ge detector (named GUOC-29) at $\Theta$ = 135$^{\circ}$ at a distance of 62 cm from the target position, used as a monitor during part of the experiment. The proton beam was stopped in a beam dump made of carbon to minimize neutron and $\gamma$-ray production. It was about 3 m downstream of the target chamber and heavily shielded with paraffin and lead to reduce neutron and $\gamma$-ray background from the beam dump. 

Neutrons in the irradiation runs were mostly produced by the  $^7$Li(p,n) reaction with the major target isotope $^7$Li (92.4\% natural isotopic abundance). The residual $^7$Be nucleus has only one bound excited state at $E_x$ = 429 keV, while higher excited states lie above the $\alpha$-$^3$He breakup threshold of $E_x$ = 1587 keV \cite{NNDC}. In the following, we label neutrons from the different $^7$Be states, including states (resonances) in the continuum above the $\alpha$-$^3$He breakup threshold with increasing excitation energy, starting with n$_0$ for the $^7$Be ground state and n$_1$ for the $E_x$ = 429 keV state.  Neutrons from continuum population, that are not identified with a known $^7$Be state will be labeled n$_X$. Neutrons can also be produced with the minor target isotope by the $^6$Li(p,n) reaction. The residual nucleus $^6$Be is particle unbound and its only relatively narrow state is the ground state, which has a (2p-$\alpha$) particle decay width of $\Delta E$ = 92 keV.  Neutrons from this state could not be identified in the experiment, we therefore considered neutrons from the $^6$Li(p,n) reaction as a continuum component. The laboratory energy threshold for the $^7$Li(p,n) reaction to the $^7$Be ground state is $E_p$ = 2.8 MeV and $E_p$ = 4.6 MeV to the $\alpha$-$^3$He breakup threshold;  for the $^6$Li(p,n) reaction it is $E_p$ = 4.3 MeV.

Irradiation runs were made at 4 different laboratory proton energies $E_p$ = 5, 6.5, 10 and 14.6 MeV. The energy range for $n_0$ and $n_1$ neutrons incident on the LaBr$_3$ detector at $\Theta$ = 45$^{\circ}$,  taking account of beam energy losses in the target foil, is $E_n$ = 2.1 - 11.8 MeV. The HP-Ge monitor detector at  $\Theta$ = 135$^{\circ}$ was present for runs with $E_p$ = 5, 6.5 and 10 MeV.   For each proton beam energy, we made a run without target to determine the beam-induced background, dominated by beam interactions within the carbon of the beam dump. Then two runs with Li targets, one with the described setup without  and a second with paraffin absorbers between target and the LaBr$_3$ and the HP-Ge detector GUOC-29 at  $\Theta$ = 135$^{\circ}$. The thickness of the paraffin absorbers was d = 30 cm for the scintillator and d = 20 cm for GUOC-29. The paraffin absorbers were very efficient for shielding the detectors from  neutrons produced at $E_p$ = 5, 6.5 and 10 MeV. At $E_p$ = 14.6 MeV, however, neutron interactions in the paraffin-shielded HP-Ge detectors became clearly visible.  Finally, GV-Orsay was removed at the same time as GUOC-29 at this energy because the radiation background in the HP-Ge detectors became too important for meaningful measurements with these detectors without time-of-flight information. 

All the detection electronics for the different detectors was based on commercially available standard NIM modules, with the exception of the module ''BaFPro'', providing neutron-$\gamma$ separation in the EDEN detector. BaFPro was made for liquid-scintillator detectors \cite{Boi} and lateron modified and specifically adapted for EDEN detectors \cite{Cav}. The BaFPro module utilizes the pulse decay-time difference between pulses induced by $\gamma$ rays and those induced by neutron interactions in the scintillator, generating a fast and total integration of the  EDEN anode signal. The photomultiplier (PM) anode signal of EDEN was split to the modified BaFPro module and a constant-fraction discriminator (CFD), the latter serving as start signal of a time-to-amplitude converter (TAC). The stop signal was the signal of the beam pulser, the TAC output thus measuring the time-of-flight between the target and the EDEN detector. Signals of the LaBr$_3$ PM anode were first processed by amplifiers before being sent to a CFD for timing and an analog-to-digital converter (ADC) for energy measurement. For the time-of-flight measurement  to  the LaBr$_3$ detector we proceeded as for EDEN, starting a TAC with the LaBr$_3$ timing signal and stopping it with the beam pulse signal. The preamplifier signals of the HP-Ge detectors were directly sent to spectroscopic amplifiers before being digitized. For these detectors, no time-of-flight signal was generated, but a timing signal was nevertheless produced for the timestamp of the acquisition system. All energy and TAC signals were digitized by ADCs of the COMET-NARVAL \cite{COMET} acquisition system of the Tandem-ALTO facility into 32768 channels. This system provides also a timestamp for each event with a resolution of 400 ps.

\subsection{Calibrations}

Energy and efficiency calibration of the LaBr$_3$ detector was done with a certified strong (384 kBq) $^{152}$Eu source placed at the target position before the neutron-irradiation runs. The resulting full-energy efficiency data in the range $E_{\gamma}$ = 122 - 1408 keV could be reproduced to better than 10\% with a Geant-3 simulation of the setup including source, detector and target chamber. The two  HP-Ge detectors employed were large-volume coaxial n-type detectors from the French-UK detector pool, formerly used in the Eurogam phase-I setup \cite{Eurogam}.  Because of the particular use of the detectors, very close to the $\gamma$-ray emitting region (GV-Orsay) or as neutron and $\gamma$-ray monitor  (GUOC-29),  the efficiency calibration has been done at different source-detector distances. Efficiency data at distances of the source from the detector front cap  $d_{sd}$ = 13.9 - 63.9 mm for GV-Orsay and $d_{sd}$ = 17.5 - 68 mm for GUOC-29, aimed at a determination of the crystal dimensions and the distance of the Ge crystal to the detector front cap, that are only partly known for Eurogam phase-I detectors. It has been done before and after the irradiation runs in a separate setup and another data acquisition system with a certified weak (6.76 kBq)  $^{152}$Eu source. The crystal geometry was then deduced with the help of Geant-3 simulations of the calibration setup, by varying the crystal geometry until the measured full-energy efficiency curves in the energy range $E_{\gamma}$ = 122 - 1408 keV were reproduced to better than 10\% at all distances $d_{sd}$. 

The intrinsic neutron detection efficiency curve of a liquid scintillator detector can be determined from a simulation of neutron interactions when the detection threshold is known \cite{Cecil}. The threshold can be determined from an en energy calibration of the scintillation detector using the full-energy peaks of low-energy $\gamma$-ray lines and the Compton edges of higher-energy lines. This method was successfully applied to EDEN and other neutron detectors based on liquid scintillators \cite{EDEN,Assie}. We proceeded in this way for the intrinsic neutron detection efficiency of our EDEN detector. Before and between the neutron irradiation runs, $\gamma$-ray spectra of $^{241}$Am, $^{22}$Na and $^{137}$Cs sources put close to the EDEN detector were obtained. The 59.5-keV full-energy peak and the Compton edges of the  511-keV, 662-keV and 1275-keV $\gamma$-ray lines in the spectra resulting from the total and/or fast-integration signals of EDEN  were then extracted by comparing with Geant-3 simulations of the $\gamma$-ray interactions in the liquid scintillator.   Additionally, the Compton edges of the 1461-keV line of $^{40}$K  in a long room-background measurement and of the 478-keV line of $^7$Li produced in the irradiation runs could also be used.   The energy calibration provided by these measurements allows a straightforward determination of the  detection threshold in terms of the electron-equivalent energy deposits in the scintillator.  The calculation of the neutron efficiency requires then neutron interaction cross sections with the scintillator components hydrogen and carbon and the knowledge of the scintillation light yield of recoiling nuclei compared to that of recoiling electrons of the same kinetic energy.

Practically, we developed a Monte-Carlo-type simulation of neutron interactions in the EDEN liquid scintillator (NE-213). It includes neutron elastic scattering off $^1$H and $^{12}$C and inelastic reactions with $^{12}$C. The latter include inelastic scattering to the 4.439-MeV state of $^{12}$C and the $^{12}$C(n,$\alpha$)$^9$Be and $^{12}$C(n,n$\alpha$)$^8$Be(2$\alpha$) reactions. Neutron-proton elastic scattering data have been taken from the NN-online site of the University of Nijmegen \cite{NN-online}, using the calculated neutron-proton scattering observables with the Nijm93 \cite{Nijm93} potential. Neutron-$^{12}$C elastic scattering data were extracted from the EXFOR  library \cite{EXFOR}, including the papers of Lane et al. \cite{Lane}, Brandenberger et al. \cite{Brand}, Demanins et al. \cite{Deman} and Haouat et al. \cite{Haouat}. Neutron-$^{12}$C inelastic reaction data were scanned from the graphs in Cecil et al. \cite{Cecil}. The scintillation light yield of recoiling protons and $\alpha$ particles has been taken from the parametrization of Cecil et al. \cite{Cecil}, except for protons below $E_p$ = 0.4 MeV where the tables of Verbinski et al. \cite{Verbinski} have been used. The light yields  of $^9$Be and $^{12}$C were taken as 1/Z$^2$ times the parameterized light yield of Jagemann et al. \cite{Jagemann} for protons. 

\begin{figure}[h]
\includegraphics[width=14 cm]{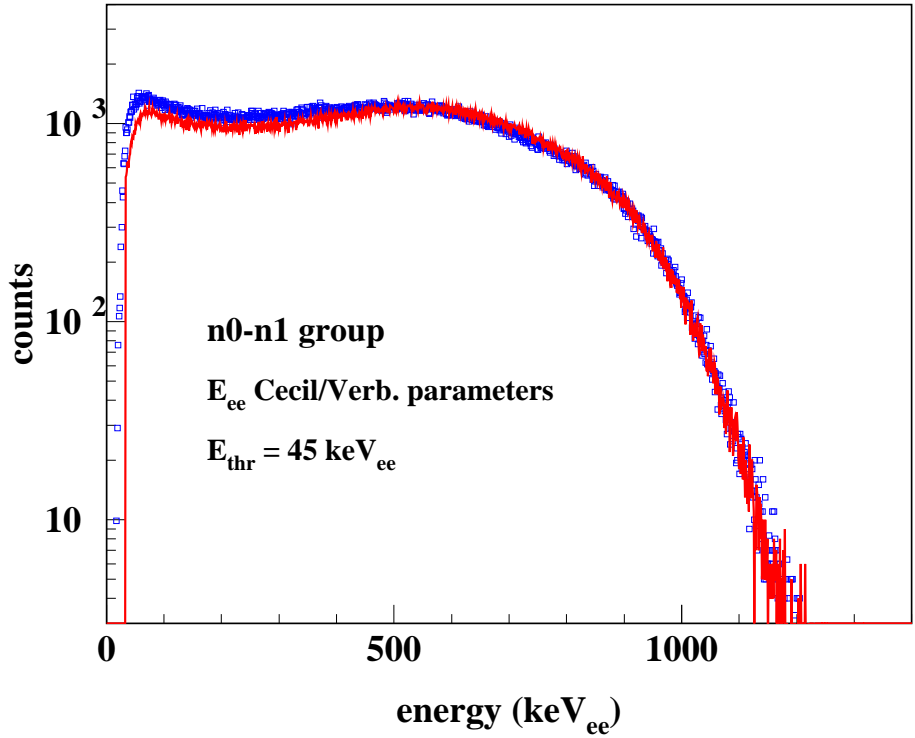}
\caption{  Light-output spectrum in electron-equivalent energy in the EDEN detector with the total-integration signal  for n$_0$-n$_1$ neutrons at $E_p$ = 5 MeV (blue squares). It has been extracted with a contour for n-$\gamma$ separation in the fast-total signal plane (see Fig. \ref{Efast_slow}) and a selection in the neutron-TOF spectrum. The solid red line shows the calculated spectrum, obtained with a Monte-Carlo type simulation of fast neutron interactions in the liquid scintillator NE-213  as described in the text.}
\label{EDENslow}
\end{figure}

These ingredients gave an excellent reproduction of the scintillator spectra for fast neutrons (an example is shown in Fig. \ref{EDENslow}), of similar quality than the results of  Ref. \cite{Zim}. They reproduce also the data of  Laurent et al. \cite{EDEN}, who measured the intrinsic neutron detection efficiency of an EDEN detector in the neutron energy range $E_n$ = 0.5 - 6 MeV. The agreement with the measured efficiency data is at the  5\%-level in the range $E_n$ = 1.5 - 6 MeV and better than 10\% down to $E_n$ = 0.75 MeV. In our setup with an electron-equivalent detection threshold of $E_{thr}$ = 45 keV$_{ee}$, this efficiency reaches 54\% at its maximum in the range $E_n$ = 0.75 - 1 MeV and decreases to 25\% at $E_n$ = 10 MeV. 

The TAC modules for the TOF determination with EDEN and the LaBr$_3$ detector were repeatedly calibrated during the irradiation runs by adding or withdrawing a passive delay of typically 16 or 32 ns on the start or stop signal. It was usually done at each new beam energy in order to place the n$_0$-n$_1$ neutron peak conveniently away from the edges of the TAC range. Their response remained relatively stable throughout the experiment with some slight fluctuations of the order of 1\%.

\section{Analysis}

Data analysis was done with the ROOT software package \cite{Root} after transforming the original  NARVAL acquisition data into ROOT-structered data, defining coincidence data for events within a time window of 8 $\mu$s. One-dimensional spectra extracted with ROOT were usually compressed to 4096 channels, except for the HP-Ge energy spectra where 8192 channels were chosen. 

\subsection{Neutron flux spectra}

\begin{figure}[h]
\includegraphics[width=14 cm]{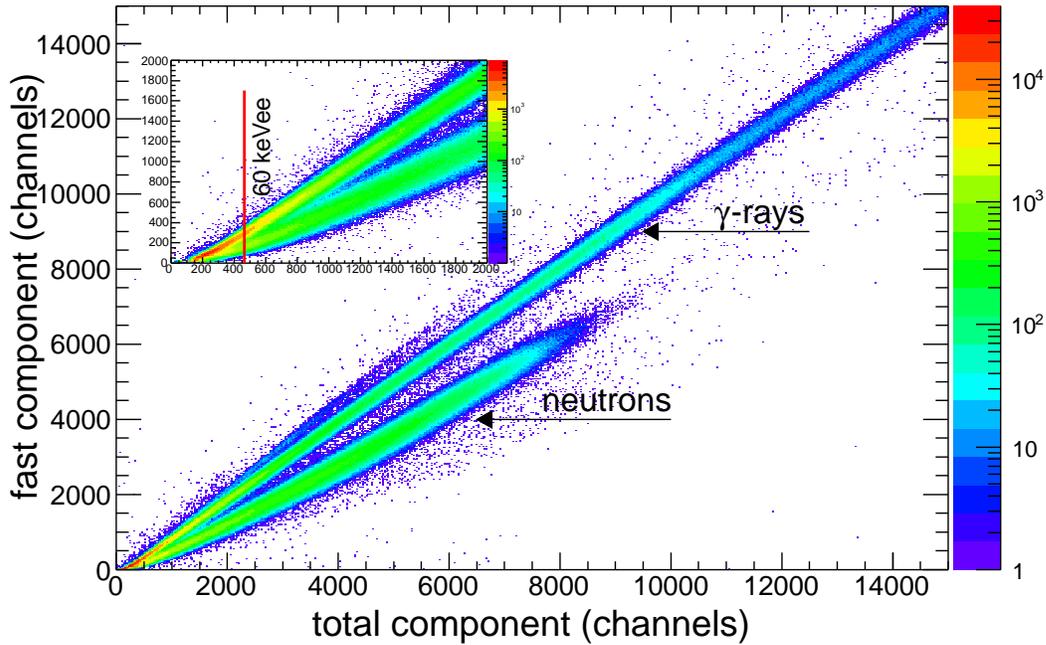}
\caption{  Scatter plot of fast- and total-integration signals of events of the EDEN detector in the neutron-irradiation run at $E_p$ = 5 MeV. The separation between the upper band from $\gamma$-ray interactions and the lower band from  neutron-induced events is excellent above the total-integration signal component $ \sim$1200 ($\sim$155 keV$_{ee}$).  For smaller signals, a good separation can still be observed down to  $\sim$450 ($\sim$60 keV$_{ee}$). }
\label{Efast_slow}
\end{figure}

\begin{figure}[h]
\includegraphics[width=14 cm]{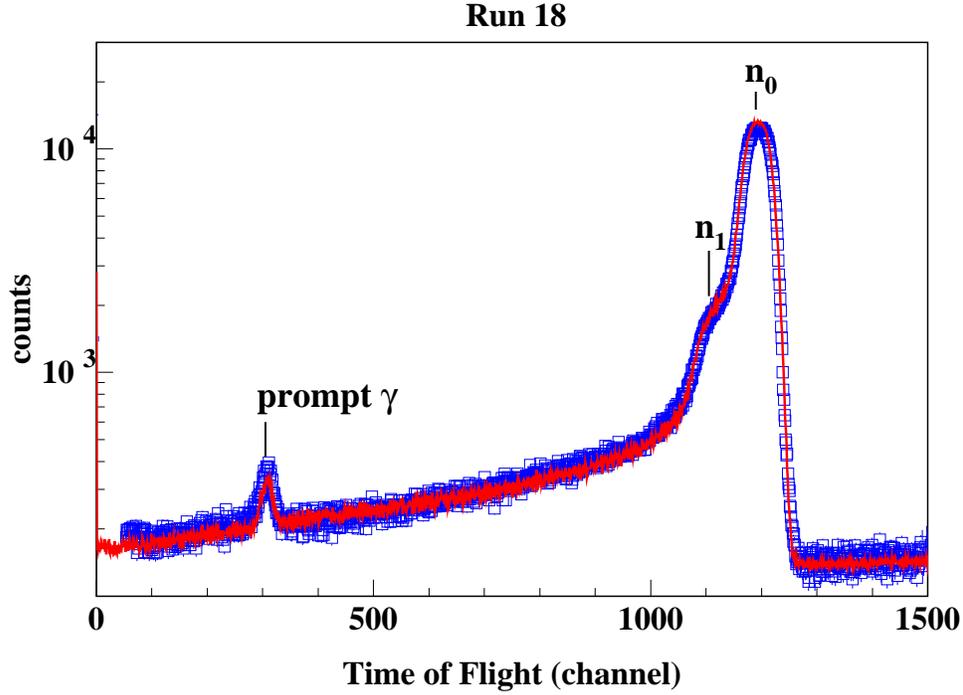}
\caption{  Neutron time-of-flight spectrum of EDEN for the irradiation run at $E_p$ = 5 MeV (open blue squares) and calculated spectrum (red line) for the best adjustment of relative intensities P(n$_0$), P(n$_1$), P(n$_X$) and two exponential components. The spectrum features also a weak prompt $\gamma$ peak at TOF channel $\sim$300 due to an incomplete n-$\gamma$ separation for very small signals. It has also been included into the Monte-Carlo type simulation that was developed for the calculation. Time of flight increases for decreasing channels. Prompt $\gamma$ emission gets its TAC stop signal in this run from an earlier beam pulse than the fast neutrons and appears therefore at the left side of the n$_0$-n$_1$ neutron peak.  }
\label{EDEN_TOF}
\end{figure}

Neutron flux spectra were deduced from the time-of-flight (TOF)  spectra of the EDEN detector, which was placed at the same angle to the beam direction as the LaBr$_3$ detector ($\Theta$ = 45$^{\circ}$). Its large distance from the target (d = 223 cm) allowed an easy separation of n$_0$ and n$_1$ neutrons from lower-energy neutrons coming from the population of higher-lying states or the continuum in $^7$Li(p,n) and $^6$Li(p,n) reactions. In a first step, contours for neutron selection in the anode signal fast and total integration plane have been defined. The neutron-$\gamma$ separation was quite effective down to  energy deposits in the liquid scintillator E $\sim$ 60 keV$_{ee}$, as can be seen in Fig. \ref{Efast_slow}. Neutron-TOF spectra for events defined in the neutron contour typically reaching down to E = 45 keV$_{ee}$ showed a very small contamination of $\gamma$-ray induced events visible by the presence of the prompt $\gamma$-ray  peak (see Fig. \ref{EDEN_TOF}). 

In order to extract the neutron-emission spectrum,  the TOF spectrum was reproduced by a Monte-Carlo type simulation of the neutron production in the target and subsequent detection in the EDEN detector. Neutron production was calculated including proton-energy loss in the target and relativistic kinematics for proton reactions with the Li isotopes. The relative weighting factors for the population of the different states in $^7$Be and, when necessary, also continuum components where taken from literature when available, otherwise they were considered free parameters. Continuum components include $^7$Li(p,n) reactions populating $^7$Be above the $\alpha$-$^3$He binding energy and the $^6$Li(p,n) reaction. The time of flight of each neutron that was inside the solid angle of the liquid-scintillator cylinder ($\Phi$ = 20 cm) of the EDEN detector was then calculated, including the flight path to an  arbitrary chosen interaction point inside the 5-cm thick scintillator. The position in the TOF spectrum was calculated based on the known position of the prompt $\gamma$-ray peak and the calibration of the TAC module. The detection efficiency of EDEN was assumed to be independent of the neutron position inside the entire solid angle spanned by the scintillator and considered zero outside, neglecting eventual border effects like in-scattering of neutrons by the structural material surrounding  the liquid scintillator. 

Fig. \ref{EDEN_TOF} shows measured and calculated neutron-TOF spectra  for an irradiation run at $E_p$ = 5 MeV. The prominent peak at EDEN\_TAC $\sim$ 1200 represents the highest-energy neutrons n$_0$ and its left-sided shoulder n$_1$ neutrons. The best adjustment of the calculation was found with relative weights P(n$_0$) = 1.0 and P(n$_1$) = 0.075, which is close to the ratio of differential cross sections d$\sigma$/d$\Omega$ at $\Theta_{cm}$ = 54$^{\circ}$ for the $^7$Li(p,n$_0$) (30.5$\pm$2.2 mb/sr) and $^7$Li(p,n$_1$) (2.98$\pm$0.21 mb/sr) reactions, obtained by an interpolation of the differential cross section data of Poppe et al. \cite{Poppe} at $E_p$ = 4.75 MeV. The width of the peaks here is essentially due to the kinematic broadening induced by the beam-energy loss in the 6 mg/cm$^2$ Li target $\Delta E_p$ = 0.42 MeV. The TOF timing resolution, including the detector resolution and beam-pulse width  (1 ns)  amounts to $\Delta$t$_{EDEN}$ = 2.7 ns, determined at the prompt $\gamma$ peak. Neutrons suffer an additional TOF broadening due to the thickness of the scintillator which adds about $\Delta$t = 2 ns for n$_0$ and n$_1$ neutrons. The energy of n$_0$ neutrons in this irradiation run at $E_p$ = 5 MeV was $E_n$ = 2.57-2.99 MeV and that of n$_1$ neutrons $E_n$ = 2.15-2.55 MeV. 

\begin{figure}[h]
\includegraphics[width=14 cm]{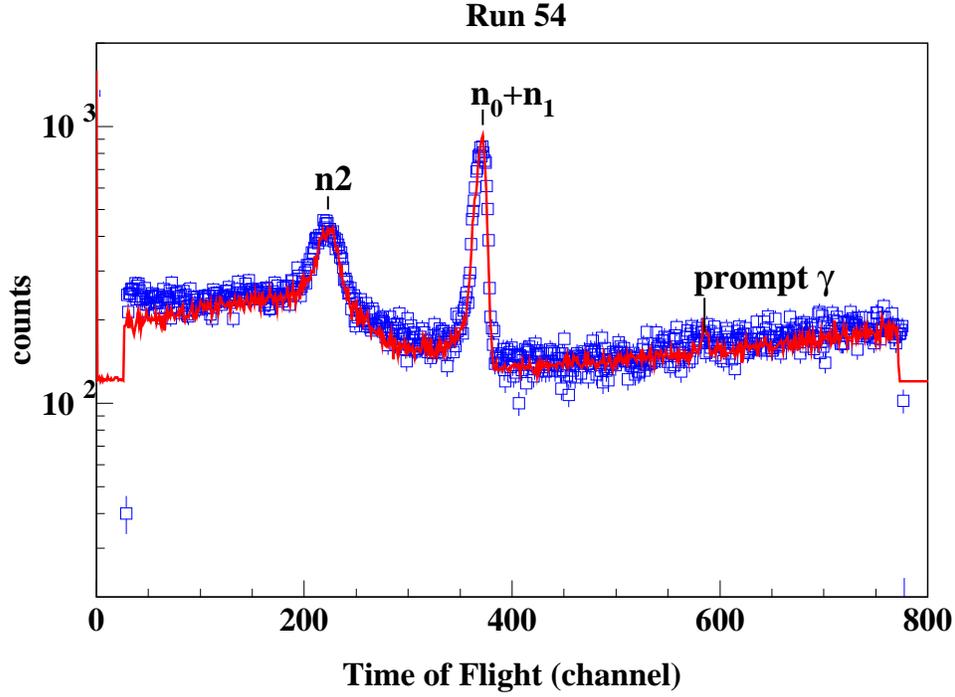}
\caption{  Neutron time-of-flight spectrum of EDEN for the irradiation run at $E_p$ = 10 MeV (open blue squares) and calculated spectrum (red line) for the best adjustment of relative intensities: P(n$_0$) = 1.0 , P(n$_1$) =  0.4, P(n$_2$) = 0.65, P(n$_3$) = 0.6, P(n$_X$) = 1.25 and two exponential components. No peak for n$_3$ is visible,  because the large width of the 3$^{rd}$ excited state in $^7$Be, $\Delta$E = 1.2 MeV \cite{NNDC}, results in a very broad energy and TOF distribution of n$_3$ neutrons with a flat maximum around channel 700. The shape of the continuum component n$_X$ has been derived from the published neutron spectrum of Poppe et al. \cite{Poppe} for the $^7$Li(p,n) reaction at $\Theta_{lab}$ = 3.5$^{\circ}$ and $E_p$ =10 MeV.  }
\label{EDEN_TOF2}
\end{figure}

The reproduction of the TOF spectra required also a constant background and two exponential tails with decay constants $\tau$ = 3-4 ns and 40-60 ns. The short tail was identified as being due to small signals whose CFD walk adjustment was not optimized.  Neutrons scattered in components like target chamber, beam line, Al table and walls and floor could account for the constant and long component, indicated by Geant-3 simulations. There could  also be a lower-energy continuum component n$_X$ from $^6$Li(p,n) and $^7$Li(p,n$^3$He$\alpha$). Both reactions are energetically possible, but their cross sections are unknown at this proton energy. These contributions are, however, clearly visible for higher proton energies together with TOF peaks of neutrons from higher excited states in $^7$Be (see Fig. \ref{EDEN_TOF2}). The determination of the n$_0$-n$_1$ neutron fluence from the EDEN-TOF spectra is estimated to be accurate to better than 10\%. This results from the quadratic sum of estimated uncertainties of 5\% for the TOF spectra analysis, 5\%  for the detector efficiency (10\% at $E_p$ = 14.6 MeV) and another 2-5\%, depending on the run, due to the contribution of the long exponential component inside the n$_0$-n$_1$ TOF range. The neutron fluence in the LaBr$_3$ detector was  calculated with the respective solid angles of the EDEN liquid scintillator and the LaBr$_3$ crystal. Fluences of n$_0$-n$_1$ neutrons ranged from 1.4$\times$10$^5$ neutrons incident on the LaBr$_3$ crystal at $E_p$ = 6.5 MeV to 2.2$\times$10$^6$ at $E_p$ = 5 MeV.

\subsection{Neutron-induced LaBr$_3$ energy spectra}

\begin{figure}[h]
\includegraphics[width=14 cm]{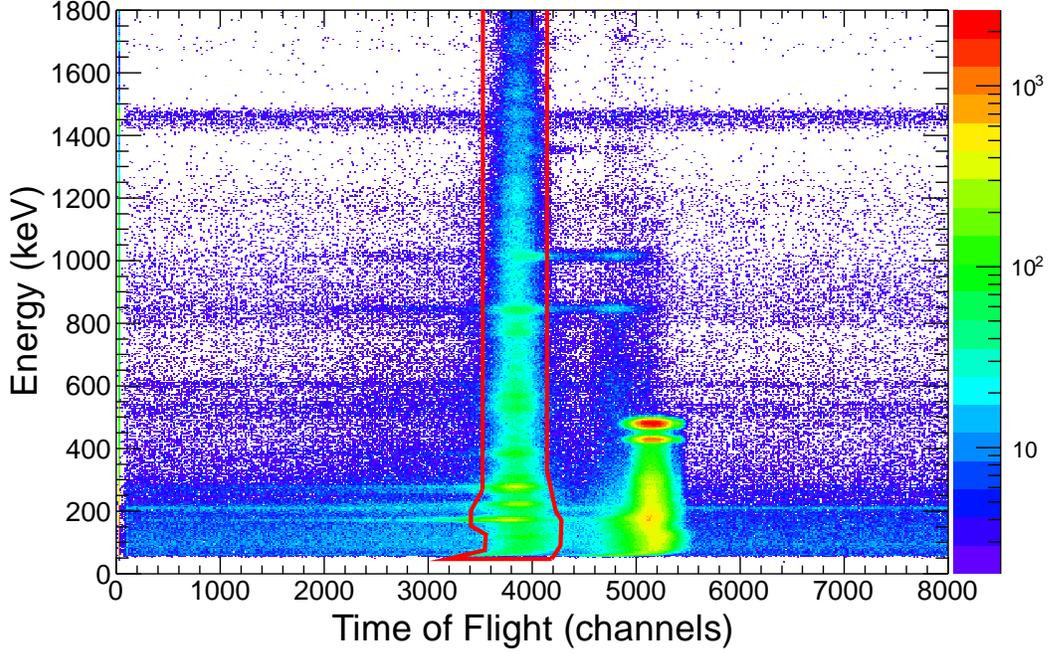}
\caption{  Part of the energy-TOF distribution of events in the LaBr$_3$ detector at $E_p$ = 5 MeV showing the prompt $\gamma$-ray band at TOF $\sim$5200 and the band from n$_0$-n$_1$-induced interactions at TOF $\sim$3900. The $\gamma$-ray band is completely dominated by the two intense  lines at $E_{\gamma}$ = 429 and 478 keV from $^7$Li(p,n$\gamma$)$^7$Be and  $^7$Li(p,p'$\gamma$)$^7$Li, respectively, and their Compton components. Two moderately strong lines at  844 keV and 1015 keV can be seen starting left  of the prompt $\gamma$ band.  They are  lines of $^{27}$Al produced by proton and neutron interactions with the target chamber and nearby material.  Several lines extending over the whole TOF range with about constant intensity are background lines like the $^{138}$La decay line at $E_\gamma$ = 1436 keV + $E_{X,Auger}$ or lines from long-lived isomeric states like the $E_x$ = 207.6 keV level of $^{79}$Br with half-life t$_{1/2}$ = 4.86 s. The red line shows the defined contour for the selection of interactions induced by n$_0$-n$_1$ neutrons.}
\label{Elabr_TOF}
\end{figure}

For the analysis of neutron-induced interactions in LaBr$_3$, only the spectra induced by n$_0$-n$_1$ neutrons were used.  To that end, it was first verified that the LaBr$_3$ TOF spectrum could be satisfactorily reproduced with the calculation as described above for the EDEN detector and the same relative weights for the different neutron components, except the exponential tails that should be different. For this purpose  the LaBr$_3$ neutron detection efficiency was obtained with a Monte-Carlo type calculation that will be described further below. The spectra were then obtained by a selection in the energy-TOF distribution (see Fig. \ref{Elabr_TOF}) around the n$_0$-n$_1$ band  guided by the EDEN and LaBr$_3$ TOF-spectra calculations. A constant radiation background, defined by the energy-TOF selection shifted to a TOF range before the prompt $\gamma$-peak was then subtracted. 

\begin{figure}[h]
\includegraphics[width=14 cm]{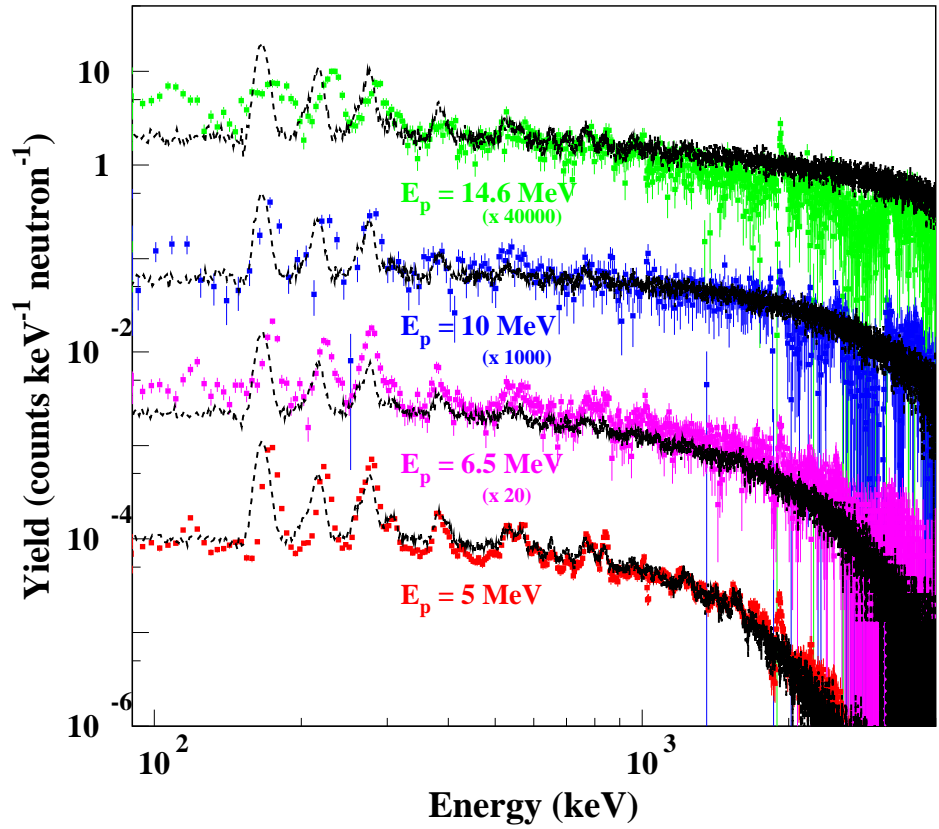}
\caption{  Symbols show neutron-induced spectra in the LaBr$_3$ detector in the irradiation runs with n$_0$-n$_1$ neutrons from the $^7$Li(p,n)-reaction at  4 different proton energies. Calculated spectra from a Monte-Carlo type simulation of neutron interactions in the scintillator are represented by the dashed lines.  }
\label{nspec}
\end{figure}

The LaBr$_3$-detector energy-TOF distribution in Fig. \ref{Elabr_TOF} shows clear evidence of a pulsed beam-related background which is dominated by proton and neutron interactions with Al-containing material. Clearly visible $\gamma$-ray lines from that component are at 844, 1015 and 2212 keV from inelastic scattering off $^{27}$Al and at $\sim$1370 keV, probably produced in $^{27}$Al(p,$\alpha$)$^{24}$Mg reactions. This beam-induced $\gamma$-ray background is also present in the runs where paraffin absorbers were placed between target and LaBr$_3$ detector. Its relative intensity in runs with and without absorbers scales approximately with the neutron flux measured in the EDEN detector. The final spectra of the LaBr$_3$-detector for n$_0$-n$_1$-neutron irradiation were thus obtained by subtracting the corresponding runs with paraffin absorber using the same energy-TOF selections as in the neutron-irradiation runs. Spectra for the 4 proton beam energies $E_p$ = 5, 6.5, 10 and 14.6 MeV are shown in Fig. \ref{nspec}. 

From these neutron-induced spectra, the total detector yield for deposited energies above 55 keV and the yield for several individual lines in the range $E$ = 166 - 308 keV were extracted. The results are listed in Table \ref{yields}, together with the mean energy of n$_0$-n$_1$ neutrons incident on the LaBr$_3$ detector. Quoted error bars on the yields include estimated uncertainties related to the extraction of n$_0$-n$_1$-induced events in the energy-TOF distribution of neutron-irradiation and background runs with paraffin absorber. They were quadratically added to uncertainties from the line-area determination and the neutron-fluence uncertainties as described above. 

In order to get cross-section estimates, the observed spectra were compared with a Monte-Carlo type modelisation of neutron interactions in the LaBr$_3$:Ce crystal. The incoming neutron flux was taken from the modelisation of the neutron-TOF spectra in the LaBr$_3$ and EDEN detectors. Each neutron was then tracked inside the crystal until either it was absorbed in a nuclear reaction or  it left the crystal boundaries. All interaction cross sections with the La, Br and Ce isotopes as well as differential cross sections for elastic neutron scattering were taken from the nuclear reaction code TALYS (version 1.4 with default parameters)  \cite{Talys}. Other ejectile and $\gamma$-ray angular distributions were taken isotropic for the sake of simplicity.   

This version of TALYS  treates at maximum the first 20 experimentally known excited states explicitly, while higher-lying states are treated implicitly in continuum bins, each bin representing an excitation energy range with specific spin distribution.  The deexcitation of those ``continuum  TALYS'' states, that are below the particle emission thresholds, was treated for the sake of simplicity as one $\gamma$-ray emission to  ''discrete TALYS'' states with probability $(2J_f + 1) \cdot  (E_i - E_f)^3 \cdot F(J_i,J_f)$, averaged over the spin distribution of the ''continuum TALYS'' state. Indices i and f  refer to emitting and receiving state, respectively, and  the factor $F(J_i,J_f)$ assures dipole transitions. For neutron capture, 3 $\gamma$-ray transitions to the ground state with random energies were generated. The energies and coordinates of the capture- and deexcitation $\gamma$ rays calculated in that code were then used as input to Geant-3 simulations of the $\gamma$-ray interactions in the LaBr$_3$:Ce crystal. More details of the modelisation will be provided in another publication \cite{modMC}.

Spectra resulting from the modelisation are shown together with the observed spectra in Fig. \ref{nspec}.  They reproduce the overall energy dependance and the line structure fairly well at the 4 energies. The total yields from the simulation agree all within the 1-$\sigma$ error bars with measured detector yields for $E$ $\geq$ 55 keV. Calculated yields agree within 1-$\sigma$ error bars for the lines  at 165.9 and 261.3 keV and underestimate the  experimental values throughout by about 1.5$\sigma$ for  the other three lines listed in Table \ref{yields}.  In the measured spectra one can observe an energy shift of several keV with respect to the nominal energy for the low-energy narrow lines, and perhaps some broadening at $E_p$ = 14.6 MeV. This energy shift can not be explained by a possible gain drift of the photomultiplier, because the energy calibration for the LaB$_3$ detector in the 200-300 keV range is well defined by decay lines of Br isomers, that are observed in spectra for the full TOF range.  It probably results from the additional scintillation light produced by recoiling nuclei after inelastic scattering, an effect that has not been included in the simulation.

In addition  to this prompt neutron-induced spectrum, intensities for lines from the deexcitation of isomeric states in the LaBr$_3$ detector material were extracted at $E_p$ = 5 and 6.5 MeV. The strongest line at 207.6 keV line is from the first excited state in $^{79}$Br: $E_x$ = 207.6 keV, 9/2$^+$, t$_{1/2}$ = 4.86 s \cite{Singh}. Lines at 260.2, 276.0 and 536.2 keV are from the second excited state in $^{81}$Br: $E_x$ = 536.2 keV, 9/2$^+$, t$_{1/2}$ = 34.6 $\mu$s \cite{Baglin}. The lifetimes of the levels in the Br isomers being much longer than the beam pulse repetition time, the line intensity was determined by a simple integration of the full TOF window. In this case  no neutron-TOF selection can be applied, meaning that together with n$_0$-n$_1$ neutrons, lower-energy neutrons from  $^6$Li(p,n) and $^7$Li(p,n) reactions and scattered neutrons also contribute  to the isomeric state population. At $E_p$ = 5 and 6.5 MeV the neutron-TOF spectra outside the n$_0$-n$_1$ peak are dominated by a constant background (see Fig. \ref{EDEN_TOF}),  that is certainly due to scattered neutrons and the afore-mentioned long exponential component, probably also mainly due to scattered neutrons. The energy spectrum of these neutrons can therefore not be deduced from the neutron-TOF spectrum.

We performed an estimation of the intensity and energy distribution of these scattered and lower-energy neutrons, based on the excellent reproduction of the measured scintillator spectra  by our Monte-Carlo type neutron-interaction simulations in the liquid scintillator NE-213 (see Fig. \ref{EDENslow}).   We compared calculated scintillator spectra for different assumed incoming neutron spectra with the measured spectra  for these events. At $E_p$ = 5 MeV, a remarkably good reproduction was found with a flat neutron spectrum for  $E_n$ = 0.2-2.5 MeV, and at  $E_p$ = 6.5 MeV with a slightly rising flux towards low energies in the range $E_n$ = 0.2-4 MeV. This component represented about 30\% at $E_p$ = 5 MeV and about 40\% at $E_p$ = 6.5 MeV in EDEN. An approximate intensity of this component in the LaBr$_3$ detector  was obtained by Geant-3 simulations of the setup with emphasis on objects that are massive or close to the target and detectors. It resulted in a scattered neutron flux that is about a factor 3 less in the LaBr$_3$ detector than in EDEN, but with very similar energy dependence and close to the energy dependence deduced from the scintillator spectra in EDEN.  The resulting yields after adding these spectra to the n$_0$-n$_1$ component are shown in Table  \ref{isomers}. For both isomeric states, the simulation agrees with the measured decay $\gamma$-ray intensities within 2-$\sigma$ error bars or to better than 30\%. These isomeric lines were also seen at the other beam energies, but a comparison with simulation was not judged worthwhile because the neutron spectra in the EDEN detector were completely dominated by a steeply rising low-energy component  $E_n$ $\leq$ 1.5 MeV,  whose intensity in the LaBr$_3$ detector was difficult to estimate. 

At the lowest proton beam energy we observed also a low-energy shoulder  at the 207.5-keV line, corresponding to a line from the second excited state in $^{19}$F: $E_x$ = 197.1 keV, 5/2$^+$, t$_{1/2}$ = 89.3 ns \cite{Tilley}. This identification is supported by its time dependence and the presence of a line close to 110 keV, corresponding to the deexcitation of the first excited state in $^{19}$F. The origin of these lines is probably neutron-inelastic scattering off F present in the teflon foil wrapped around the LaBr$_3$ crystal.

At $E_p$ = 5 MeV, several lines emitted from the LaBr$_3$ crystal could be seen in the energy spectrum of the close-by  Ge detector (GV-Orsay). Their intensities are presented in Table \ref{GVO}  and compared with the result of a  simulation of the LaBr$_3$ $\gamma$-ray emission. The detection efficiency of GV-Orsay for  a $\gamma$-ray emitted within the LaBr$_3$ crystal was done with a Geant-3 simulation of the detection setup. It was the same simulation as described above for neutron irradiation of the LaBr$_3$:Ce crystal, except that  the detector GV-Orsay was added to the Geant-3 simulation. The Ge crystal dimensions in the simulations were taken from the efficiency calibrations with the $^{152}$Eu source as described above in the section about calibrations. We estimate an uncertainty of less than 10\% for the Ge full-energy efficiency in this configuration, including an effect of multiple hits in the Ge crystal from $\gamma$-ray cascades that amounts to about 5\%. This uncertainty was included into the quoted error bars of the emission yields and quadratically added to uncertainties from the line intensity determination and  neutron flux. The agreement between simulation and experiment is reasonable, below the 1.5-$\sigma$ level for the 6 lines. 

\subsection{Neutron-induced Ge energy spectra}

At the two lowest proton beam energies $E_p$ = 5 and 6.5 MeV, neutron-induced spectra in the HP-Ge monitor detector GUOC-29, placed at $\Theta_{lab}$ = 135$^{\circ}$ could be extracted and analyzed. The corresponding neutron energies at this angle are below 3 MeV, where few states in the different Ge isotopes can be excited. For the majority of them, in particular the most important, neutron inelastic scattering cross sections have been measured in the required neutron-energy range. Neutron spectra incident on the Ge crystal and the crystal dimensions are known, meaning that practically all important parameters for the calculation of neutron-induced spectra are given. These data could therefore be used for an independent test of the neutron-interaction  program described above for the LaBr$_3$ crystal. 

The neutron spectrum incident on the Ge crystal of the GUOC-29 detector has been calculated with differential cross sections for n$_0$ and n$_1$ neutrons at $\Theta_{lab}$ = 135$^{\circ}$ from Poppe et al. \cite{Poppe}. Flux normalization was done taking the n$_0$-n$_1$ neutron flux determined with EDEN, modulated by the ratios of differential cross sections d$\sigma_n$(135$^{\circ}$)/d$\sigma_n$(45$^{\circ}$) and solid angles d$\Omega$(GUOC-29)/d$\Omega$(EDEN). Because TOF information is not available for GUOC-29, the detector spectra contain also interactions of the low-energy and scattered neutron component. This component was obtained by Geant-3 simulations of the setup as described above for the isomeric states in LaBr$_3$ irradation. It resulted in similar neutron spectra at the positions of EDEN and GUOC-29 with, however, about a factor of 3 less scattered neutrons with respect to n$_0$-n$_1$ neutrons in GUOC-29 than in EDEN. Corresponding neutron flux spectra  were added to n$_0$-n$_1$ neutrons, respresenting about 10\% and 15\% of the total flux at $E_p$ = 5 and 6.5 MeV, respectively.  

\begin{figure}[h]
\includegraphics[width=14 cm]{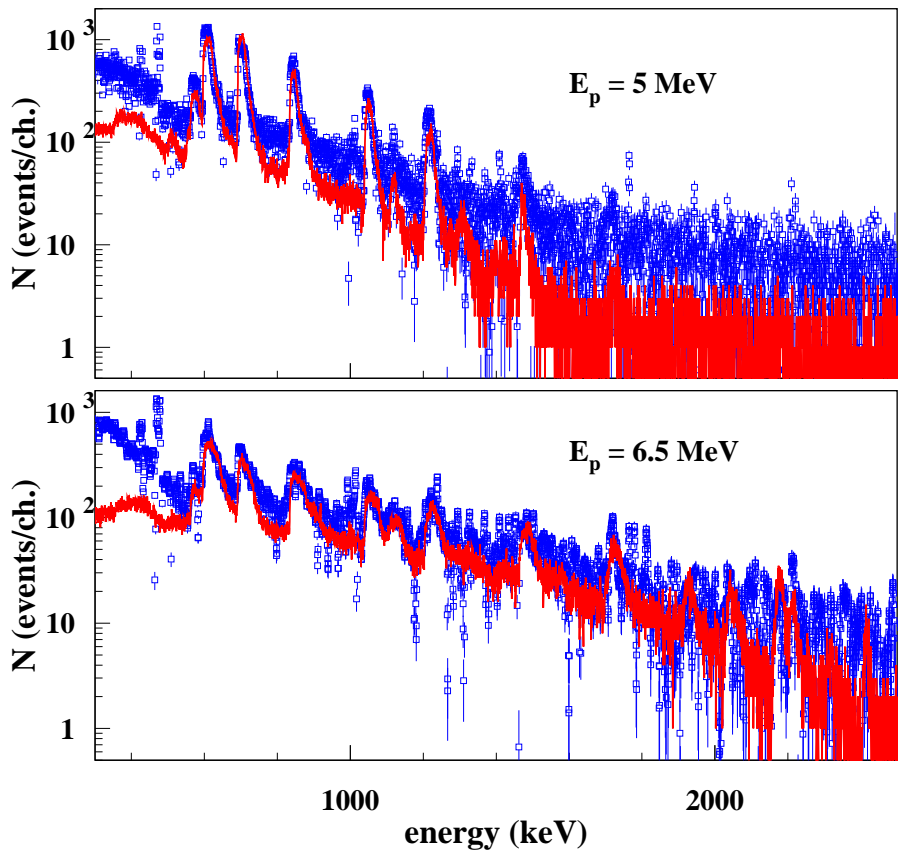}
\caption{  Blue symbols show the neutron-induced spectra in the HP-Ge detector at $\Theta$ = 135$^{\circ}$ in the irradiation runs with  neutrons from the $^7$Li(p,n) reaction at  $E_p$ = 5 MeV and 6.5 MeV. The calculated spectra from a Monte-Carlo type simulation of neutron interactions in the Ge crystal are represented by the red lines (see text). }
\label{GeSpec}
\end{figure}

There are also background $\gamma$-ray lines in the spectra of GUOC-29, in particular the proton-induced 429-keV and 478-keV lines from reactions with $^7$Li, lines from proton and neutron interactions with Al and Fe and the 511-keV line and the 2223.2-keV  line from neutron capture on H. This $\gamma$-ray background was subtracted with the help of Geant-3 simulations of the setup. The resulting spectra at $E_p$ = 5 and 6.5 MeV are shown in Fig. \ref{GeSpec} together with  results of  the Monte-Carlo type simulation of neutron interactions in the Ge crystal. This simulation was realized  with neutron inelastic scattering cross sections  taken from Table V in Chung et al. \cite{Chung}.  Measured values were used when available in that table, otherwise the calculated values were taken. For extrapolations to lower energies, the excitation functions of Fig. 4 in Ref. \cite{Chung} were used. For excited levels of Ge isotopes not listed in that table, cross sections have been taken from TALYS \cite{Talys}. For the energy deposit signal of Ge recoils, the prescription of Ref. \cite{LewSmi} was used. More details about the simulations will be presented in another publication \cite{modMC}. The calculation reproduces nicely the triangular features typical for (n,n'$\gamma$) reactions in Ge detectors. The  continuum component  below 500 keV, underestimated by the simulation, is very probably due to $\gamma$-ray background not subtracted by the Geant-3 simulations, e.g. neutron capture $\gamma$ rays produced in the detector,  in Al and Fe structures and in the concrete walls and floor. 

Measured and calculated yields for the strongest $\gamma$-ray lines are listed in Table \ref{GeLines}.  The calculations reproduce the measured yields generally well within the 1-$\sigma$ error bars, with the notable exception of the 689.6-keV line of $^{72}$Ge at $E_p$ = 5 MeV. The latter corresponds to a 0$^+$- 0$^+$ transition emitting a conversion electron that has not been measured in the experiment of Chung et al. \cite{Chung}. Our results imply a cross section of $\sim$180 mb at $\overline{E}_n$ =  1.6 MeV, lower by  $\sim$30\% with respect to the theoretical value in their Table VI for $E_n$ = 1.75 MeV.  These results validate our approach for neutron-interaction simulations and shows the ability of the developed Monte-Carlo type program to predict the neutron response of $\gamma$-ray detectors.

\subsection{Calculated yields for fast neutrons in LaBr$_3$ detectors}
 
In order to facilitate the estimation of background  in LaBr$_3$ detectors induced by fast neutrons, based on the good agreement so far with measured data, we present calculated yields for prompt and delayed emission and for some individual lines in LaBr$_3$  crystals of 4 different sizes. The explored energy range is $E_n$ = 0.5 - 10 MeV, for neutrons incident on  cylindrical crystals with diameters and lengths of 2.54, 3.81, 5.08 and 7.62 cm (1, 1.5, 2 and 3 inch). For the calculation,  the geometry of the present experimental setup was chosen, i.e. the detector front faces are at 40 cm from the neutron emission origin. Nuclear reaction cross sections are the same as above for the comparison of measured and calculated yields and spectra. In the neutron energy range covered by the experiment ($E_n$ = $\sim$2-12 MeV), estimated uncertainties for these calculations are 15\% for the total prompt emission yields and typically 30\% for the individual lines (see Table \ref{yields}). Outside this range an additional uncertainty may be applied for possible shortcomings of TALYS in cross-section energy dependencies, that should not, however, exceed 30\%. 

\begin{figure}[h]
\includegraphics[width=17 cm]{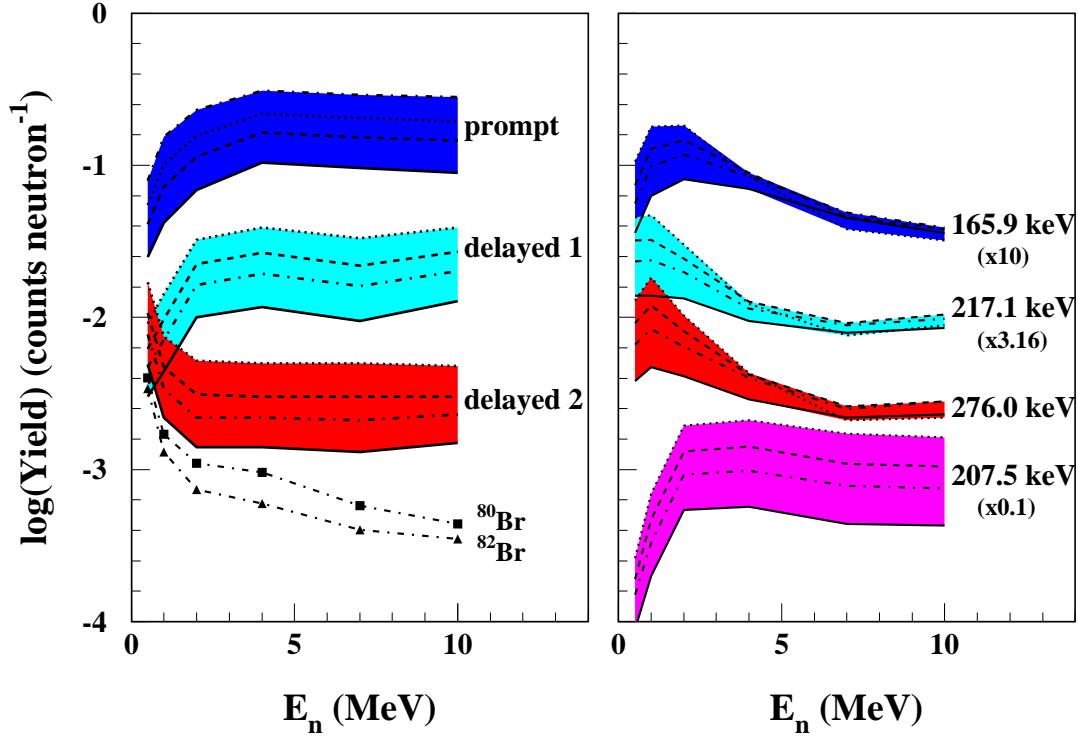}
\caption{  Calculated yields for neutron-induced emission in cylindrical LaBr$_3$ crystals of sizes (diameter and length) 2.54, 3.81, 5.08 and 7.62 cm (inside the shaded regions: full, dot-dashed, dashed and dotted lines, respectively). Left panel: total yields for energy deposits $E$ $\geq$ 20 keV for prompt emission, short-lived activity (``delayed 1'',  50 ns $\leq$ $t_{1/2}$ $\leq$10 m), long-lived activity (``delayed 2'',  $t_{1/2}$ $>$10 m). Calculated yields for the activity of 2 Br radioisotopes  in a 3.81-cm LaBr$_3$ crystal are also shown.  Right panel: Calculated yields for 3 prompt emission lines and the delayed line at 207.5 keV from $^{79}$Br$^m$. The yields of the 165.9, 217.1, and 207.5-keV lines have been multiplied by different factors for better visibility.}
\label{CalYields}
\end{figure}

\begin{figure}[h]
\includegraphics[width=17 cm]{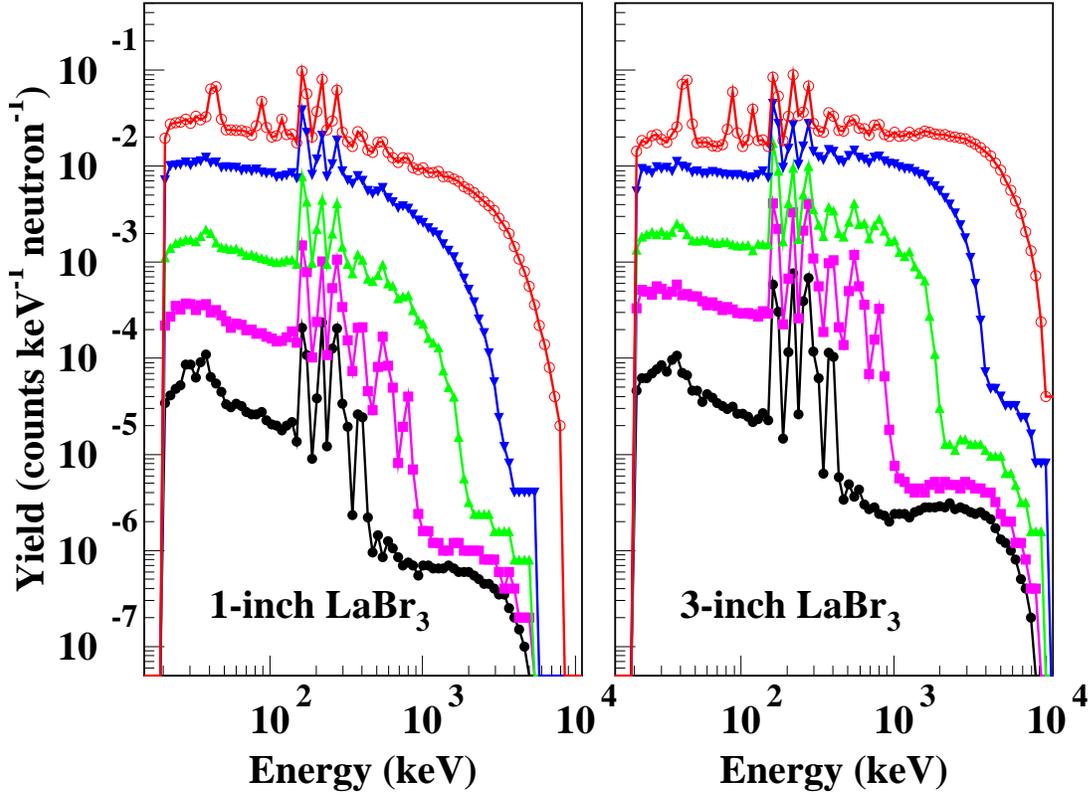}
\caption{  Calculated spectra  of neutron-induced interactions in LaBr$_3$ detectors with cylindrical crystals of diameter and length 2.54 cm (left) and 7.62 cm (right). Incoming neutron energies are (from bottom to top) $E_n$ = 0.5 MeV (filled black circles), 1 MeV (magenta squares; Yield x 4), 2 MeV (green triangles; Yield x 16), 4 MeV (blue triangles; Yield x 80) and 10 MeV (red open circles; Yield x 400). Only prompt $\gamma$-ray emission (t$_{1/2}$ $\leq$ 10 ns) is considered, delayed $\gamma$-ray emission and the energy deposit of charged reaction products and recoiling nuclei is neglected in the calculation. Yields are simulated counts per incoming neutron and keV.}
\label{SynSpec}
\end{figure}

The results of the calculations are summarized in Fig. \ref{CalYields}. The neutron-induced emission is dominated by the prompt emission, the three strongest prompt emission lines (165.9, 217.1 and 276.0 keV) are shown in the right panel. Delayed emission due to short-lived activity (50 ns $\leq$ $t_{1/2}$ $\leq$ 10 m) contributes typically 10-15\%, and is dominated by the isomers  $^{79}$Br$^m$ (t$_{1/2}$ = 4.86 s) and $^{81}$Br$^m$ (t$_{1/2}$ = 34.6 $\mu$s), except at $E_n$ = 0.5 MeV, where the isomer $^{82}$Br$^m$ (t$_{1/2}$ = 6.13 m) dominates. The yield of the strongest delayed-emission line at 207.5 keV from  $^{79}$Br$^m$ is shown in the right panel of Fig. \ref{CalYields}. The long-lived activity (t$_{1/2}$ $\geq$ 10 m) is only significant at $E_n$ = 0.5 MeV, dominated by $^{80}$Br (t$_{1/2}$ = 17.68 m) and $^{82}$Br (t$_{1/2}$ = 35.282 h), produced by neutron capture. The calculation for all activities assumes equilibrium between radioisotope production and decay. Calculated spectra for prompt emission at $E_n$ = 0.5, 1, 2, 4 and 10 MeV for crystal sizes of 2.54 and 7.62 cm are displayed in Fig. \ref{SynSpec}.

\section{Conclusion}

We have determined neutron-interaction yields in a 1.5-inch cylindrical LaBr$_3$:Ce crystal  in the energy range $\overline{E}_n$ =  2.8 - 11.6 MeV. Total yields for energy deposits $E \geq$ 55 keV in the crystal and yields for 5 individual $\gamma$-ray lines have been extracted at 4 different neutron energies. Yields of lines from the decay of the isomers of $^{79}$Br and $^{81}$Br have been obtained at the two lowest neutron energies. We have developed  a Monte-Carlo type simulation of neutron interactions in the LaBr$_3$:Ce crystal with cross sections from the nuclear reaction code TALYS that reproduces total yields to better than 15\%, well within 1-$\sigma$ uncertainties. Yields of individual lines are reproduced within typically 30\% uncertainty. We have then at our disposal a powerful tool for the prediction of neutron-induced background in LaBr$_3$ crystals with arbitrary dimensions  in e.g. space radiation environments and nuclear physics experiments at particle accelerators. For this purpose, calculated yields and spectra for several neutron energies and crystals of different sizes are presented. Conversely, thanks to the present results, one can use the LaBr$_3$ detectors to measure neutron fluxes and spectral distributions in nuclear physics experiments. 

In addition, neutron-interaction yields in a cylindrical HP-Ge crystal have been obtained at $\overline{E}_n$ =  1.6 and  2.4 MeV. The Monte-Carlo type simulation with neutron inelastic scattering cross sections from literature showed very good agreement with yields for 8 individual lines. This validates again the developed simulation program and shows its ability to predict the neutron response of $\gamma$-ray detectors. It validates also implicitely the theoretical cross sections for some low-lying states of $^{70}$Ge and $^{72}$Ge, in particular for the 689.6-keV line where no measured values are available. Furthermore, a good prescription for interactions of fast neutrons and the scintillation light yields of nuclear recoils in the liquid scintillator NE-213 was found. It gives an excellent description of measured neutron-induced scintillator spectra down to electron-equivalent energy deposits of 45 keV and reproduces accurately the measured intrinsic detection efficiency for neutrons in the range $E_n$ = 0.5 - 6 MeV.

\newpage.


\begin{table}

\caption{Total yields for prompt emission above $E\simeq$ 55 keV and individual  yields for 5 $\gamma$-ray lines of  the 1.5-inch LaBr$_3$:Ce detector in the irradiations with neutrons of average energy $\overline{E}_n$ at proton beam energy $E_p$. The lines are from transitions of the first excited states in $^{139}$La  (5/2$^+$ 165.9 $\rightarrow$ 7/2$^+$ g.s.) and $^{81}$Br (5/2$^-$ 276.0 $\rightarrow$ 3/2$^-$ g.s.), and the 2$^{nd}$ to 4$^{th}$ excited states of  $^{79}$Br above the $E_x$ = 207.6 keV isomer (5/2$^-$ 217.1 $\rightarrow$ 3/2$^-$ g.s.); (3/2$^-$ 261.3 $\rightarrow$ 3/2$^-$ g.s.); (1/2$^-$,3/2$^-$ 306.5 $\rightarrow$ 3/2$^-$ g.s.). Yields are expressed in [counts $\cdot$ neutron$^{-1}$], numbers in parenthesis are the uncertainties on the last digits. The values in the 2$^{nd}$, 4$^{th}$, 6$^{th}$ and 8$^{th}$ row are the results of the neutron interaction simulation. Proton beam energies $E_p$ and mean neutron energies $\overline{E}_n$ are in MeV.}
\vspace{0.5 cm}

\begin{tabular}{lllcccccc}
 & $E_p$ & $\overline{E}_n$   & $E \geq $ 55 keV  & 165.9 keV  & 217.1 keV &  261.3 keV &  276.0 keV &   306.5 keV  \\
 \hline \\
 Exp. & 5  & 2.8 & 0.129(13)  & 9.9(13)$\cdot$10$^{-3}$   & 6.5(7)$\cdot$10$^{-3}$ & 2.6(7)$\cdot$10$^{-3}$ & 9.8(25)$\cdot$10$^{-3}$ & 1.6(6)$\cdot$10$^{-3}$\\
 Sim. & & & 0.130  & 1.2$\cdot$10$^{-2}$   & 5.2$\cdot$10$^{-3}$ & 2.1$\cdot$10$^{-3}$ & 5.8$\cdot$10$^{-3}$ & 1.2$\cdot$10$^{-3}$\\
 Exp. & 6.5 & 4.2 & 0.137(20) & 6.3(15)$\cdot$10$^{-3}$   & 5.4(13)$\cdot$10$^{-3}$   & 2.1(17)$\cdot$10$^{-3}$  & 9.0(35)$\cdot$10$^{-3}$   & 1.5(6)$\cdot$10$^{-3}$   \\
Sim. & &   & 0.155 & 7.6$\cdot$10$^{-3}$   & 3.2$\cdot$10$^{-3}$   & 1.1$\cdot$10$^{-3}$  & 3.6$\cdot$10$^{-3}$   & 5.3$\cdot$10$^{-4}$   \\
 Exp. & 10 & 7.4 & 0.160(31) & 5.2(19)$\cdot$10$^{-3}$   & 3.2(12)$\cdot$10$^{-3}$   & 7.5(38)$\cdot$10$^{-4}$   & 4.3(10)$\cdot$10$^{-3}$   & 1.1(6)$\cdot$10$^{-3}$   \\ 
 Sim. & & & 0.145 & 4.8$\cdot$10$^{-3}$   & 2.3$\cdot$10$^{-3}$   & 7.0$\cdot$10$^{-4}$   & 2.6$\cdot$10$^{-3}$   & 4.0$\cdot$10$^{-4}$   \\ 
 Exp. & 14.6 & 11.6 & 0.125(45) &  3.4(14)$\cdot$10$^{-3}$   & 3.4(14)$\cdot$10$^{-3}$   & -&  $\leq$ 4.6$\cdot$10$^{-3}$ & $\leq$ 1.0$\cdot$10$^{-3}$\\
 Sim. & &    & 0.141 &  4.8$\cdot$10$^{-3}$   & 2.6$\cdot$10$^{-3}$   & 8.7$\cdot$10$^{-4}$&  2.9$\cdot$10$^{-3}$ & 4.0$\cdot$10$^{-4}$\\
 
\end{tabular} 
\label{yields}
\end{table}


\begin{table}
\caption{Results of $\gamma$-ray line emission from  isomeric states in Br isotopes in the irradiation with neutrons of mean energies $\overline{E}_n$. $\overline{E}_n$ is lower than in Table \ref{yields}, because it takes into account scattered and lower energy neutrons (see text). Yields ($E_n$ $\geq$ 0.2 MeV) for the 207.5-keV line from the first excited state in $^{79}$Br (9/2$^+$ 207.6 $\rightarrow$ 3/2$^-$ g.s.), the two transition lines at 260.2 keV of $^{81}$Br (9/2$^+$ 536.2 $\rightarrow$ 5/2$^-$ 276.0) and 276.0 keV of $^{81}$Br (5/2$^-$ 276.0 $\rightarrow$ 3/2$^-$ g.s.), and the sum line of the deexcitation cascade at 536.2 keV (536.2 $\rightarrow$ 276.0 $\rightarrow$ g.s.) The values in the 2$^{nd}$ and 4$^{th}$ row are the results of the neutron interaction simulation. Definition of yields, parenthesis and particle energies: see Table \ref{yields}.}
\vspace{0.5 cm}

\begin{tabular}{lcccccc}
  & $E_p$ & $\overline{E}_n$  & 207.5 keV  & 260.2 keV  & 276.0 keV &  536.2 keV \\
 \hline \\
Exp. & 5 & 2.6 & 7.3(12)$\cdot$10$^{-3}$   & 1.3(3)$\cdot$10$^{-3}$ & 1.1(2)$\cdot$10$^{-4}$ & 1.7(3)$\cdot$10$^{-3}$ \\
Sim. & &  &  9.6$\cdot$10$^{-3}$ & 1.6$\cdot$10$^{-3}$ & 1.4$\cdot$10$^{-3}$ & 2.2$\cdot$10$^{-3}$ \\
Exp. & 6.5 & 3.4 & 8.0(25)$\cdot$10$^{-3}$  & - & - & 2.0(9)$\cdot$10$^{-3}$  \\
Sim  & &  & 8.7$\cdot$10$^{-3}$ & 1.5$\cdot$10$^{-3}$ & 1.4$\cdot$10$^{-3}$ & 2.3$\cdot$10$^{-3}$ \\
\end{tabular} 
\label{isomers}
\end{table}

\begin{table}
\caption{Results of $\gamma$-ray line emission of the LaBr$_3$:Ce crystal, induced by interactions of neutrons with mean energy $\overline{E}_n$ = 2.4 MeV. Counts per incoming neutron for 6 lines determined from the spectrum of a large volume HP-Ge detector close to the LaBr$_3$:Ce crystal (1$^{st}$ row) and simulation (2$^{nd}$ row). The lines are from transitions of low-lying levels in the Br isotopes: 207.6, 217.1, 260.2, 261.3 and 276.0 keV (see Tables \ref{yields},\ref{isomers});   381.5 keV ($^{79}$Br; 5/2$^+$ 381.5 $\rightarrow$ 3/2$^-$ g.s.); 538.2 keV  ($^{81}$Br; 1/2$^-$3/2$^-$ 538.2 $\rightarrow$ 3/2$^-$ g.s.). Numbers in parenthesis are the uncertainties on the last digits.}
\vspace{0.5 cm}

\begin{tabular}{lcccccc}
 & 207.5 keV &  217.1 keV   & 260.2+261.3 keV  & 276.0 keV &  381.5 keV & 538.2 keV \\
 \hline \\
Exp. & 1.2(6)$\cdot$10$^{-2}$ & 1.2(4)$\cdot$10$^{-2}$   & 9.8(38)$\cdot$10$^{-3}$ & 2.9(6)$\cdot$10$^{-2}$ & 1.2(3)$\cdot$10$^{-2}$ & 7.6(19)$\cdot$10$^{-3}$  \\
Sim.  & 1.45$\cdot$10$^{-2}$ & 1.74$\cdot$10$^{-2}$   & 1.54$\cdot$10$^{-2}$ & 3.7$\cdot$10$^{-2}$ & 1.01$\cdot$10$^{-2}$ & 7.9$\cdot$10$^{-3}$\\
\end{tabular}  
\label{GVO}
\end{table}

\begin{table}
\caption{Results of $\gamma$-ray line emission induced in a Ge crystal of 7 cm diameter and 8 cm length by interactions of neutrons with mean energy $\overline{E}_n$ = 1.6 MeV at $E_p$ = 5 MeV and $\overline{E}_n$ = 2.4 MeV at $E_p$ = 6.5 MeV. Counts per incoming neutron Y$_{exp}$ for 8  lines determined from the triangular features typical for (n,n'$\gamma$)-reactions in Ge detectors and calculated values Y$_{sim}$ from the Monte-Carlo type simulation of neutron interactions  in the Ge crystal. Two structures are composed of several lines, not resolved in the measured spectra. In the calculation, the $\sim$600-keV structure has a 4.8\% contribution from the 608.35-keV line, the $\sim$1210-keV structure has a 33\% contribution from the 1215.5-keV line at $E_p$ = 5 MeV. These contributions are 10\% and 24\%, respectively, at $E_p$ = 6.5 MeV. Numbers in parenthesis are the uncertainties on the last digits.}
\vspace{0.5 cm}

\begin{tabular}{llllll}
 & & $E_p$ = 5 MeV & & $E_p$ = 6.5 MeV & \\
  level $E_x$ (keV) & $E_{line}$ (keV) & Y$_{exp}$  &  Y$_{sim}$ &  Y$_{exp}$  &  Y$_{sim}$\\
 \hline \\
  $^{76}$Ge 563.92 & 563.92 & 2.8(6)$\cdot$10$^{-3}$ &  2.9$\cdot$10$^{-3}$ & 4.0(29)$\cdot$10$^{-3}$  & 3.8$\cdot$10$^{-3}$ \\
 $^{74}$Ge 595.85  & 595.85 & 1.38(25)$\cdot$10$^{-2}$ &  1.44$\cdot$10$^{-2}$ & 1.7(4)$\cdot$10$^{-2}$ & 1.6$\cdot$10$^{-2}$\\
 + $^{74}$Ge 1204.21  & 608.35  &  &   &  &  \\
 $^{72}$Ge 691.43  & 689.6 & 9.4(16)$\cdot$10$^{-3}$ &  1.37$\cdot$10$^{-2}$ & 1.0(3)$\cdot$10$^{-2}$  & 9.6$\cdot$10$^{-3}$ \\
 $^{72}$Ge 834.01  & 834.01 & 5.0(10)$\cdot$10$^{-3}$ &  5.3$\cdot$10$^{-3}$ & 7.5(25)$\cdot$10$^{-3}$  & 6.7$\cdot$10$^{-3}$ \\
 $^{74}$Ge 1463.76  & 867.90 & 4.7(20)$\cdot$10$^{-4}$ &  3.3$\cdot$10$^{-4}$ & -  &   8.6$\cdot$10$^{-4}$\\
 $^{70}$Ge 1039.49  & 1039.49 & 2.7(5)$\cdot$10$^{-3}$ &  2.7$\cdot$10$^{-3}$ & 4.3(13)$\cdot$10$^{-3}$  & 4.4$\cdot$10$^{-3}$ \\
 $^{76}$Ge 1108.44  & 1108.41 & 5.0(20)$\cdot$10$^{-4}$ &  2.8$\cdot$10$^{-4}$ & - & 9.1$\cdot$10$^{-4}$ \\
 $^{74}$Ge 1204.21  & 1204.21 & 1.72(32)$\cdot$10$^{-3}$ &  1.51$\cdot$10$^{-3}$ & 3.6(12)$\cdot$10$^{-3}$  & 3.1$\cdot$10$^{-3}$ \\
  + $^{70}$Ge 1215.54  & 1215.5  &  &   &  &  \\

\end{tabular} 
\label{GeLines}
\end{table}

\end{document}